\documentclass[aps,prb,reprint,floatfix,superscriptaddress]{revtex4-1}
\usepackage[english]{babel}                                                           
\usepackage[utf8]{inputenc}                                                            
\usepackage{bbold}
\usepackage{epsfig}
\usepackage{amsmath}
\usepackage{units}
\usepackage[colorlinks]{hyperref}
\usepackage{enumerate}
\usepackage[justification=raggedright]{caption}	
\usepackage{color}		  				

\usepackage{lipsum}

\begin{document}

\title{Engineering of Chern insulators and circuits of topological edge states}
\author{Emma L. Minarelli}
\affiliation{School of Physics, University College Dublin, Belfield, Dublin 4, Ireland}
\affiliation{Institute for Theoretical Physics and Center for Extreme Matter and Emergent Phenomena,
Utrecht University, Leuvenlaan 4, 3584 CE Utrecht, The Netherlands}
\author{Kim P\"oyh\"onen}
\affiliation{Department of Applied Physics, Aalto University, P.~O.~Box 15100,
FI-00076 AALTO, Finland}
\author{Gerwin A. R. van Dalum}
\affiliation{Institute for Theoretical Physics and Center for Extreme Matter and Emergent Phenomena,
Utrecht University, Leuvenlaan 4, 3584 CE Utrecht, The Netherlands}
\author{Teemu Ojanen}
\affiliation{Department of Applied Physics, Aalto University, P.~O.~Box 15100,
FI-00076 AALTO, Finland}
\affiliation{Laboratory of Physics, Tampere University of Technology, Tampere FI-33101, Finland}
\author{Lars Fritz}
\affiliation{Institute for Theoretical Physics and Center for Extreme Matter and Emergent Phenomena,
Utrecht University, Leuvenlaan 4, 3584 CE Utrecht, The Netherlands}

\begin{abstract}
Impurities embedded in electronic systems induce bound states which under certain circumstances can hybridize and lead to impurity bands.
Doping of insulators with impurities has been identified as a promising route towards engineering electronic topological states of matter. In this paper we show how to realize tuneable Chern insulators starting from a three dimensional topological insulator whose surface is gapped and intentionally doped with magnetic impurities.  The main advantage of the protocol is that it is robust, and in particular not very sensitive to the impurity configuration. We explicitly demonstrate this for a square lattice of impurities as well as a random lattice. In both cases we show that it is possible to change the Chern number of the system by one through manipulating its topological state. We also discuss how this can be used to engineer circuits of edge channels.
\end{abstract}
\maketitle

\section{Introduction}

Electronic topological states of matter have been at the forefront of condensed matter research for more than thirty years. The starting point was the discovery of the quantum Hall state and the subsequent identification of the topological origin of the associated quantization of the transverse conductivity. More recently it has become apparent that this state of quantum matter fits into a bigger scheme of states, now called topological insulators \cite{hasankane,qi}. This whole development was largely triggered by the prediction and subsequent observation of the quantum spin Hall insulator. Since then, many more systems have entered the picture. On a theoretical level, it has been shown that the knowledge of only a small number of elementary symmetries (time-reversal, parity, and chirality) and the dimensionality allows prediction of whether a system might in principle possess a non-trivial topological invariant~\cite{schnyder2008,kitaev2009,Qi2008,schnyderNJP2010,bernevigbook}. This scheme can be extended if lattice symmetries and statistical effects are taken into account~\cite{fu2011,slager2013,chiu2013,morimoto2013}. A further strategy which has been applied to generate topological states is dissipation and/or driving. All these theoretical insights and ideas have lead to a spur of experimental activities with one of the main questions being whether one can tailor-make a specific kind of non-trivial insulator. This goal is pursued in a variety of ways, ranging from cold atom physics to more traditional condensed-matter setups.  
 
A variant of this engineering is intentional 'pollution': the starting point is an insulator on whose surface adatoms are deposited in a regular manner. These adatoms or 'impurities' bind electronic in-gap states which hybridize, leading to impurity bands with potentially topological character \cite{choy,np,brau1,klin,vazifeh,pientka2,poyh,heimes,reis,bry,west,balents,kimme}. A recent example is the deposition of ferromagnetic atoms in a chain on superconductors. These magnetic impurities individually bind a pair of Shiba bound states. Once these bound states hybridize, one obtains artificial wires which have been shown to potentially host topological bands. This was, for instance, demonstrated in an experiment where iron atoms were arranged in a chain structure on a superconductor. These efforts have lead to the observation of Majorana-like signatures at the ends of the chain \cite{np2,feldman,ruby,pawlak}. In a related experiment \cite{ruby2} using Co atoms these modes were absent meaning that the protocol does not seem robust and independent of details. Theoretically this scheme has also been extended to higher dimensions \cite{nakosai}, for instance leading to two-dimensional topological superconductors with a wide range of notably high Chern numbers\cite{ront1,ront2,kaladzhyan}.

\begin{figure}[h]
\centering
\includegraphics[width=0.25\textwidth]{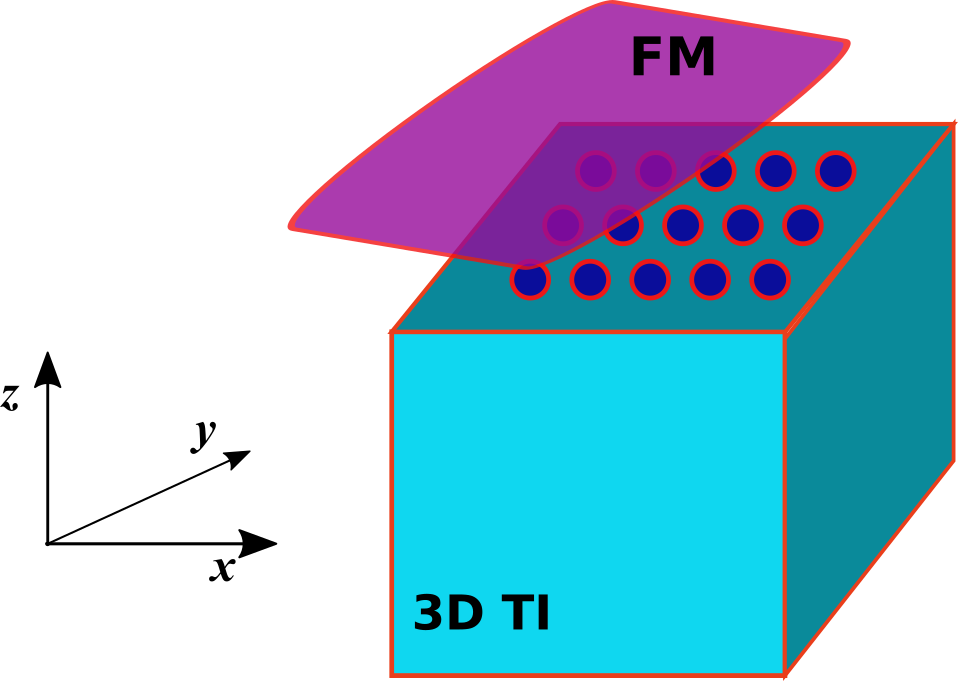}
\includegraphics[width=0.15\textwidth]{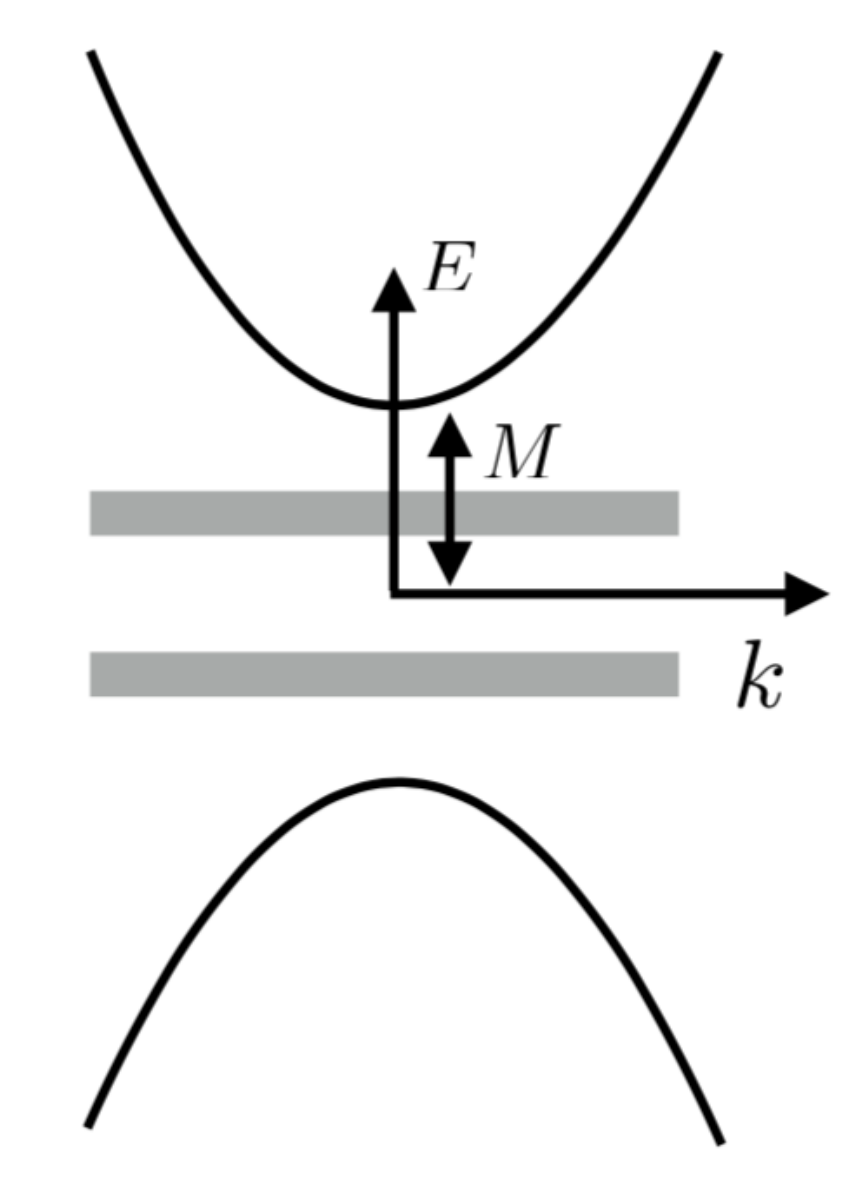}
\caption{a): Sketch of the setting for a square lattice implementation: a three-dimensional strong topological insulator (3D TI) serves as  base. On its top surface, the vicinity of a ferromagnetic material (FM) induces an energy gap. A lattice of impurities, with mixed scalar and magnetic potential, is deposited on the surface. The engineered setting acts as a non-trivial Chern insulator. b): Spectrum of a magnetic topological insulator surface with a mass gap M. Hybridized impurities give rise to subgap bands (grey) carrying non-zero Chern numbers. }\label{fig:sketch}
\end{figure}

In this paper we follow in this spirit and show that depositing a lattice of magnetic impurities on the gapped surface of a three dimensional topological insulator (this can be achieved by proximity to a ferromagnet) leads to a non-trivial Chern insulator, see Fig.~\ref{fig:sketch}. The virtue of our construction is that the Chern number can be locally tuned by impurities, and that it is generic and independent of the precise details of the deposition.  

The paper is organized in the following way: we start with a discussion of the effective model system and introduce the key quantities in Sec.~\ref{sec:model}. We then discuss the physics of impurities deposited on the surface of the topological insulator in Sec.~\ref{sec:impurity}, where we first discuss a single impurity and then move on to a lattice of impurities. In Sec.~\ref{sec:impband}, we discuss the impurity in-gap band and its topological properties in different lattice configurations.  In Sec.~\ref{sec:engset} we explain the implications of our proposal to the fabrication of topological circuits, and we summarize our findings in Sec.~\ref{sec:conclusion}.

\section{Model and key quantities}\label{sec:model}

We consider the top surface of a 3D strong topological insulator \cite{hasankane} in the vicinity of a ferromagnetic layer breaking time-reversal symmetry, see Fig.~\ref{fig:sketch}. An appropriate minimal model for the surface is given by the massive Dirac Hamiltonian
\begin{equation}
 \hat{H}_S(\textbf{k}) =v_{F} (k_{x}\tau_{x}+k_{y}\tau_{y}) + M\tau_{z} - \mu \mathbb{1}\;,
\end{equation}
with $v_{F} $ being the Fermi velocity, $\mu$ the chemical potential, and $\tau_{i}$ with $i=x,y,z$ the Pauli matrices. The energy spectrum is given by $E_{\pm}({\textbf{k}})=\pm \sqrt{v_F^2 k^2+M^2}-\mu$ with a gap size $\Delta_{gap}=2|M|$, which is induced due to the proximity to the ferromagnet. For the remainder of this paper we will set the chemical potential $\mu =0$.

The fundamental object in our study is the retarded Green function $\hat{\mathcal{G}}^{+}_{0}(E)$:
\begin{equation}
\hat{\mathcal{G}}^{+}_{0}(E) =   \hat{\mathcal{G}}_{0}(E+ i 0^{+})\;,
\end{equation}
which in real space obeys the differential equation
\begin{equation}\label{eq:de}
((E+i 0^{+})\mathbb{1}- \hat{H}_S)\hat{\mathcal{G}}^{+}_{0}(\textbf{r},\textbf{r\ensuremath{'}};E)=\delta(\textbf{r}-\textbf{r\ensuremath{'}})\;,
\end{equation}
where $\textbf{r}= (x,y)^{T}$ is a 2D surface vector. The solution to Eq.~\eqref{eq:de} is translationally invariant and reads
\begin{widetext} 
 \begin{equation}\label{eq:FullGreen}
\hat{\mathcal{G}}^{+}_{0}(\textbf{r}_{i},\textbf{r}_{j};E)=\hat{\mathcal{G}}^{+}_{0}(\textbf{r}_{ij};E) =- \dfrac{1}{2 \pi v_F^2} \Big( (E \mathbb{1}+M \tau_{z})K_{0}\bigg(\dfrac{|\textbf{r}_{ij}|}{\xi}\bigg) 
 -i \dfrac{v_{F}}{\xi}  \dfrac{x_{ij} \tau_{x} + y_{ij} \tau_{y}}{|\textbf{r}_{ij}|} K_{1}\bigg(\dfrac{|\textbf{r}_{ij}|}{\xi}\bigg)  \Big)\;,
 \end{equation}
\end{widetext}
where $\textbf{r}_{ij}=\textbf{r}_{j}-\textbf{r}_{i}\neq 0$ is the relative coordinate, $\xi$ is the correlation length given by $\xi = v_{F} / \sqrt{M^{2}- E^{2}}$, and $K_{z}$ are the $z$th order modified Bessel functions of the second kind. A full derivation of Eq.~\eqref{eq:FullGreen} can be found in the appendix. We are only interested in in-gap phenomena, meaning $|E|<|M|$ in the remainder of the paper.

Next, we place static, structureless impurities on the surface, described by the Hamiltonian
\begin{equation}
 \hat{H}_I(\textbf{r})= \sum_{i=1}^{N} (V_{0} \mathbb{1}+V_{M}\tau_{z}) \delta(\textbf{r}-\textbf{r}_{i})\;,
 \end{equation}
where $N$ is the number of impurities and ${\textbf{r}_i}$ with $i=1,...,N$ denotes their respective positions.
We assume that an impurity is strictly local and has both a scalar ($V_{0}$) and a magnetic ($V_{M} $) contribution (again, this should strictly be understood as a minimal model).

As shown in the appendix, the bound state solutions of the full Hamiltonian can be found as the solutions to the integral equation~\cite{mahan} 
\begin{equation}\label{eq:inteq}
\psi(\textbf{r}) = \int \mathrm{d}^{2}\textbf{r\ensuremath{'}} \hat{\mathcal{G}}^{+}_{0}(\textbf{r} - \textbf{r\ensuremath{'}};E) \hat{H}_I(\textbf{r\ensuremath{'}}) \psi( \textbf{r\ensuremath{'}})\;.
\end{equation}

\section{Impurity Physics}\label{sec:impurity}

In the following we will first study a single impurity and the associated bound states and then move to multiple impurities. 
\subsection{Single Impurity Model}
The bound states of the system with one impurity on the TI surface which (without loss of generality) is located at position ${\textbf{r}}=0$ are given by the solutions of the equation
\begin{equation}\label{Eq:SingleEq}
\psi(0) = \hat{\mathcal{G}}^{+}_{0}(0;E)(V_{0} \mathbb{1}+V_{M}\tau_{z})\psi(0)\;,
\end{equation} 
where
\begin{equation}\label{Eq:SingleGreen}
\begin{aligned}
\hat{\mathcal{G}}^{+}_{0}(0;E) & =  - \dfrac{1}{4\pi v_F^2} ( E \mathbb{1} + M \tau_{3}) \log\left(1+\dfrac{D_{bulk}^2}{M^2-E^2}\right)\\
& = \begin{pmatrix}
\mathcal{G}_{0}(0;E) & 0 \\
0 & -\mathcal{G}_{0}(0;-E)
\end{pmatrix} 
\end{aligned}
\end{equation}
is the local Green function, with  $D_{bulk} \gg |M|$ being the bulk bandwidth (see again the appendix). Note that the imaginary part of this Green function is zero in the region of interest.

\begin{figure}[t]
\centering
\includegraphics[width=0.45\textwidth]{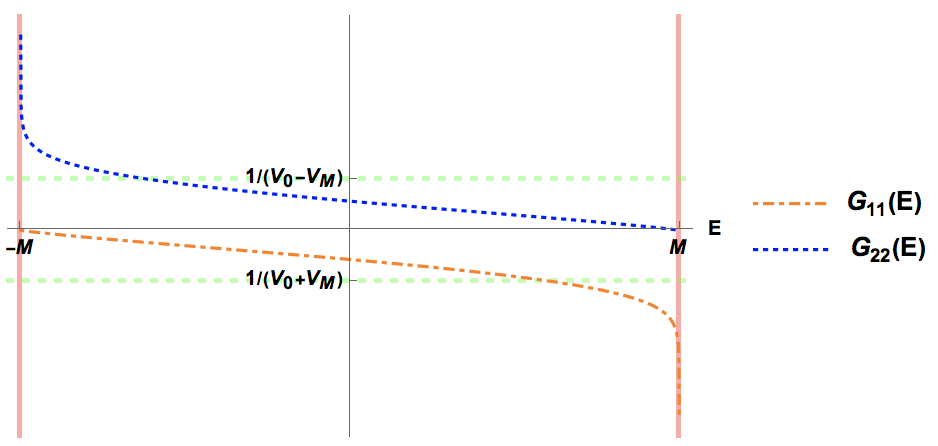}
\caption{The single-impurity Green function matrix elements are diverging and finite at opposite gap boundaries $|M|$ (thick line). At the intersection among these curves and the impurity potential (dashed), we find the energy of the corresponding bound state. For reference we have taken $M>0$ in the above figure.}\label{fig:GreenTrends}
\end{figure}

Non-trivial bound state solutions of Eq.~\eqref{Eq:SingleEq} exist for the energies which solve
\begin{equation}\label{eq:bs}
 \det \Big( \mathbb{1} - \hat{\mathcal{G}}^{+}_{0}(0;E)(V_{0} \mathbb{1}+V_{M}\tau_{z})\Big) = 0 \;.
\end{equation} 
The local Green function is diagonal and the diagonal elements are both related to the same function $\mathcal{G}_{0}(0;E)$. This function has the property that its real part approaches zero as $E \to -M$ and it diverges logarithmically as $E \to M$. This implies that one can graphically construct the non-trivial solutions to Eq.~\eqref{eq:bs} by intersecting the diagonal element $G_{11}=\mathcal{G}_{0}(0;E)$ with the line defined by $1/(V_0+V_M)$ and $G_{22}=-\mathcal{G}_{0}(0;-E)$ with the line defined by $1/(V_0-V_M)$. Let us first consider $M>0$, see Fig.~\ref{fig:GreenTrends}. Since $G_{11}$ is now strictly negative, $1/(V_0+V_M)$ must be negative as well and $1/(V_0-V_M)$ must be positive to intersect with $G_{22}$. Consequently, we can have zero to two bound states, depending on the potential scatterer: (i) For $V_M>0$ we have either zero ($|V_0|<V_M$) or one bound state ($V_0>V_M$ or $V_0<-V_M$). (ii) For $V_M<0$ we either have one bound state ($V_0<V_M$ or $V_0>|V_M|$) or two bound states ($|V_0|<|V_M|$). Since we are interested in the formation of two in-gap bound states, for the remainder we take $V_0=0$ and $V_{M}<0$. Under the assumption that $V_M$ is chosen such that $|E_{BS}|\ll |M|$, it follows that the energy of the two in-gap bound states is given by
\begin{equation}
\begin{split}
E_{BS} &= \mp  M \mp {\dfrac{4\pi v_F^2}{V_M}\log^{-1}\left(1+\dfrac{D_{bulk}^2}{M^2-E_{BS}^2}\right)}\\
&\approx  \mp  M \mp {\dfrac{2\pi v_F^2}{V_M}\log^{-1}\left(\dfrac{D_{bulk}}{|M|}\right)}\;.
\end{split}
\end{equation}
Going through the same process for $M<0$ gives the exact same bound state energies, but requires $V_M>0$ instead. The above expression is therefore valid for any $M\neq 0$ and $V_0=0$, as long as the sign of $V_M$ is opposite to the sign of $M$.

\subsection{Multiple Impurities}
We consider a number of impurities on the TI surface. In isolation, each of them possesses two non-degenerate energy levels for $V_0=0$ and $V_M<0$ or $V_M>0$, depending on the sign of $M$. Bringing these impurities closer together leads to overlap between neighboring bound state wave functions causing the bound states to disperse. 
Eq.~\eqref{eq:inteq} for multiple impurities reads
\begin{equation}\label{eq:mi}
\begin{split}
\psi(\textbf{r}^{i})=&\, V_{M} \hat{\mathcal{G}}^{+}_{0}(0;E)\tau_{z}\psi(\textbf{r}_{i})\\
+& \sum_{\textbf{r}_{j} \neq \textbf{r}_{i}} V_{M} \hat{\mathcal{G}}^{+}_{0}(\textbf{r}_{ij};E)\tau_{z}\psi(\textbf{r}_{j})\;, \end{split}
\end{equation}
where we have explicitly split into the local bound state contribution and the overlap with adjacent impurities. Eq.~\eqref{eq:mi} is exact and in the following we  map it onto a tight-binding model. The first term on the right-hand side corresponds to the isolated impurity and the corresponding bound states whereas the second term describes the coupling to the other bound states in the system. This overlap is set by the Green function, whose range is controlled by the correlation length $\xi$. For $|\textbf{r}| \gg \xi$ it asymptotically behaves as $\hat{\mathcal{G}}^{+}_{0}(\textbf{r};E)\propto e^{-|\textbf{r}| / \xi}/\sqrt{|\textbf{r}| / \xi}$. In the limit where the mean distance between impurities $\bar{a}$ is equal or larger than the correlation length, {\it i.e.}, $\bar{a}\gtrsim \xi$ we are in the dilute limit. In this dilute limit the overlap is exponentially suppressed, implying that the impurity bound state energies are only weakly dispersed. The bandwidth of the impurity bands is therefore expected to be much smaller than the gap size $2|M|$, and all of the bound state energies remain of the order of $E_{BS}$. Below we use this to approximate the dilute limit of Eq.~\eqref{eq:mi} as a tight-binding problem of hybridized bound states hopping on a lattice.
Our starting point is the atomic limit in which the local energy is given by the isolated impurity bound states and we assume that the overlap between adjacent bound states is small, {\it i.e.}, $E \sim E_{BS}\simeq 0 $. This results in the following effective integral equation:
\begin{equation}\label{eq:gftb}
\left( \mathbb{1} - V_{M} \hat{\mathcal{G}}^{+}_{0}(0;E) \tau_{z} \right)\psi(\textbf{r}_{i}) \approx  V_{M}\!\sum_{\textbf{r}_{j} \neq \textbf{r}_{i}}  \hat{\mathcal{G}}^{+}_{0}(\textbf{r}_{ij};0)\tau_{z}\psi(\textbf{r}_{j})\;.
\end{equation}
Expanding the left-hand side to linear order in $E$, Eq.~\eqref{eq:gftb} is equivalent to a tight-binding model. The effective Hamiltonian that satisfies the eigenvalue equation is given by
\begin{align}\label{eq:ham}
E \psi(\textbf{r}_{i}) =& E_{BS}\tau_{z} \psi(\textbf{r}_{i}) +2\pi v_F^2 g  \tau_z\!\sum_{\textbf{r}_{j} \neq \textbf{r}_{i}} \hat{\mathcal{G}}^{+}_{0}(\textbf{r}_{ij};0) \tau_z\psi(\textbf{r}_{j})\notag\\
=&\sum_{\textbf{r}_{j}} \hat{H}_{ij} \psi(\textbf{r}_{j})\;,
\end{align}
where $g=2\log^{-1} \left( 1+\dfrac{D_{bulk}^2}{M^2}\right) \approx \log^{-1} \left( \dfrac{D_{bulk}}{|M|}\right) \ll 1$. We see from Eq.~\eqref{eq:ham} that the bandwidth of the impurity bands is controlled by $g$ thus indeed small. This is consistent with the assumption that $E\sim E_{BS}$, therefore justifying the approximations. 
In explicit terms, the matrix elements of the tight-binding Hamiltonian read 
\begin{equation}\label{Eq:HEff}
\hat{H}_{ij}  =
\begin{pmatrix}
h_{\textbf{r}_{ij}} & \Delta_{\textbf{r}_{ij}}\\ \Delta_{-\textbf{r}_{ij}}^{\star} & -h_{\textbf{r}_{ij}}
\end{pmatrix}
\end{equation}
where
\begin{equation*}
h_{\textbf{r}_{ij}} =
\begin{cases}
E_{BS} \quad  , \quad \textbf{r}_{i}= \textbf{r}_{j} \\
-g M K_{0}\bigg(\dfrac{|\textbf{r}_{ij}|}{\xi_{0}}\bigg) \quad  , \quad  \textbf{r}_{i} \neq \textbf{r}_{j} 
\end{cases}
\end{equation*}
\begin{equation*}
\Delta_{\textbf{r}_{ij}} =
\begin{cases}
0  \quad  , \quad \textbf{r}_{i}= \textbf{r}_{j} \\
-ig\dfrac{v_F}{\xi_0} \dfrac{x_{ij} - i y_{ij}}{|\textbf{r}_{ij}|} K_{1}\bigg(\dfrac{|\textbf{r}_{ij}|}{\xi_{0}}\bigg) \quad  , \quad  \textbf{r}_{i} \neq \textbf{r}_{j} 
\end{cases}
\end{equation*}
and $\xi_0=v_F/|M|$.

\section{Lattice implementation in 2D}\label{sec:impband}

In the following we first discuss an arrangement of the impurities on a square lattice and then move to a random lattice. We explicitly show that the result of our study does not depend on any details of the arrangement of impurities.
\subsection{Square lattice}
We consider a 2D system with time-reversal symmetry breaking, thus the in-gap band structure can be classified by the first Chern number, {\it i.e.}, the topological invariant {\it C} \cite{bernevigbook,haldane}. A convenient way to evaluate \textit{C} for our system is to rewrite Eq.~\eqref{Eq:HEff} as a  Bloch Hamiltonian
\begin{equation}
\hat{H}_{\textbf{k}} = \sum_{\textbf{r}} e^{i \textbf{k} \cdot \textbf{r}} \hat{H}(\textbf{r})\;,
\end{equation}
where the summation runs over all lattice sites. It is straightforward to decompose the Bloch Hamiltonian into the basis of Pauli matrices $\boldsymbol{ \tau}$:
\begin{equation}
\hat{H}_{\textbf{k}}  = \underbrace{\Re(\Delta_{\textbf{k}})}_{d_{x}(\textbf{k})} \cdot \tau_{x} \underbrace{-\Im(\Delta_{\textbf{k}})}_{d_{y}(\textbf{k})} \cdot \tau_{y} +  \underbrace{h_{\textbf{k}}}_{d_{z}(\textbf{k})}  \cdot \tau_{z} = \textbf{d}(\textbf{k}) \cdot \boldsymbol{ \tau}\;.
\end{equation}

The normalised $\textbf{d}(\textbf{k})$-vector $\widehat{\textbf{d}}(\textbf{k}) = \dfrac{\textbf{d}(\textbf{k})}{|\textbf{d}(\textbf{k})|}$ defines the mapping between the 2D Brillouin zone and the surface enclosed by the $\widehat{\textbf{d}}(\textbf{k})$-vector on a unit Bloch sphere. Setting the impurity lattice constant to unity, we calculate the integer-valued topological invariant $C$ with the formula
\begin{equation} 
C = \pm \dfrac{1}{4 \pi} \int\limits_{-\pi}^{\pi} dk_{x}\! \int\limits_{-\pi}^{\pi} dk_{y} \widehat{\textbf{d}}(\textbf{k}) \cdot\Big(\dfrac{ \partial}{\partial k_{x}} \widehat{\textbf{d}}(\textbf{k}) \times \dfrac{ \partial}{\partial k_{y}} \widehat{\textbf{d}}(\textbf{k})\Big)\;,
\end{equation}
counting the number of times the $\widehat{\textbf{d}}(\textbf{k})$-vector wraps around the unit Bloch sphere.

\begin{figure}[t]
\centering
\includegraphics[width=0.95\linewidth]{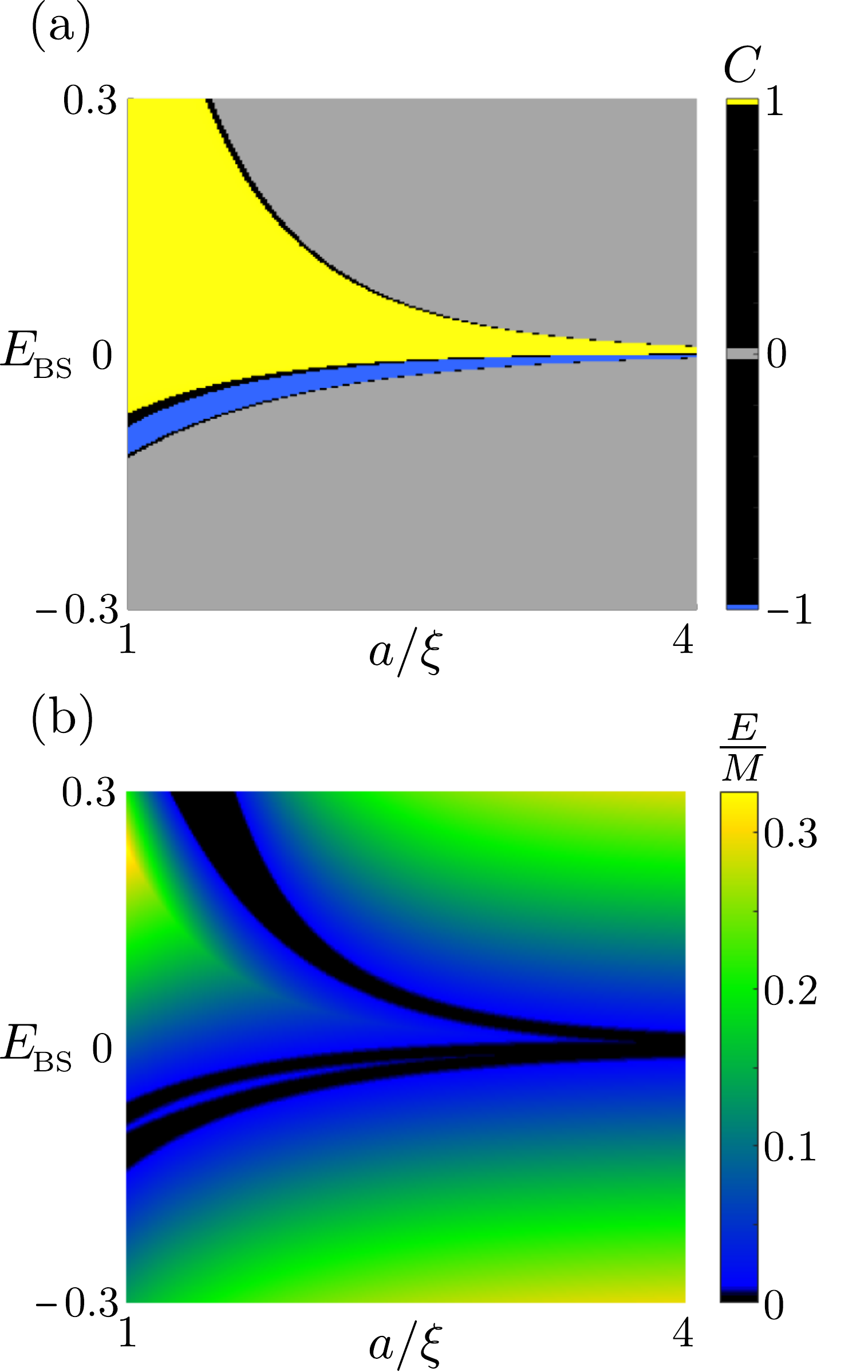}
\caption{a): Topological phase diagram of a square impurity lattice. In the considered regime the system is dominated by the Chern number $C=1$ state.  b): The energy gap diagram corresponding to the phase diagram in Fig.~a). While the $C=1$ phase may exhibit a robust gap, the narrow $C=-1$ strip is nearly gapless and thus has little practical relevance.   Calculated using $g = 0.15$.}\label{fig:square}
\end{figure}

\begin{figure}[h]
\centering
\includegraphics[width=0.8\linewidth]{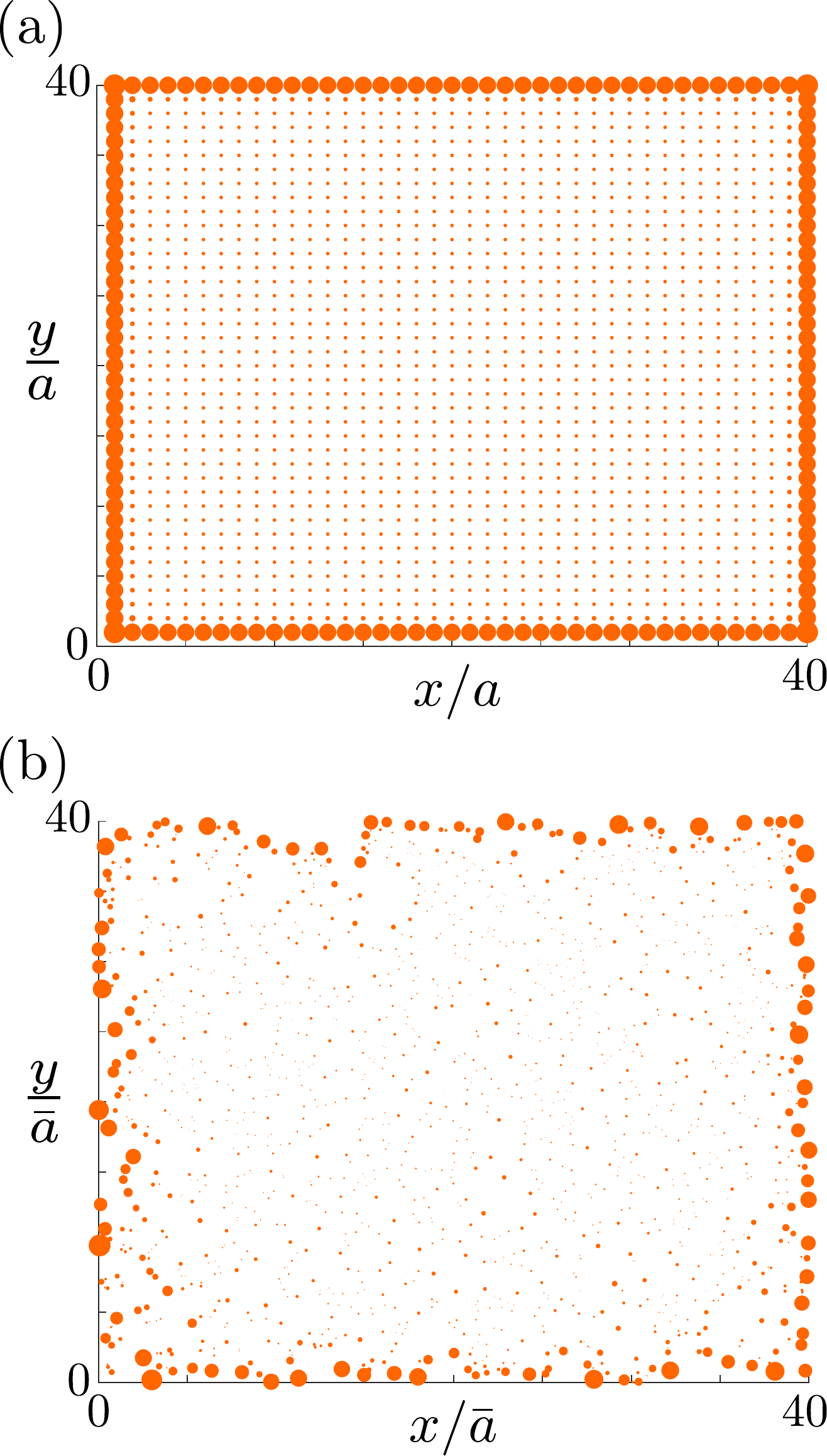}
\caption{a): Subgap zero-energy local density of states (LDOS)  for a square lattice with open boundaries for $E=0$ reveals the existence of topological edge modes. Calculated for a 40$\times$40 square lattice with $g = 0.15$, $a/\xi = 2$, $E_{BS} = 0.05$, using a Lorentzian broadening of the Dirac delta function b): Subgap LDOS for $E=0$ for a random lattice with open boundaries. Calculated for a system of 1600 particles with $g = 0.15$, $\bar a/\xi = 1.4$ and $E_{BS} = 0.05$. }\label{fig:ldos}
\end{figure}

In the case of a square lattice, the outcome is that either $C=0$ (corresponding to the trivial phase), or $|C|=1$, indicating a topological phase. The topological phase diagram as a function of relevant system parameters is shown in Fig.~\ref{fig:square}. In the deep-dilute regime captured by our model, the non-trivial parameter regime of the phase diagram is dominated by the $C=1$ phase. Note that flipping the sign of the magnetization would invert the sign of the Chern number. As a consequence of the non-trivial topology, systems with boundaries exhibit topologically protected edge states at subgap energies.  We have diagonalized the model  \eqref{Eq:HEff} on a finite lattice and plotted the local density of states (LDOS) $\rho(E,\mathbf{r}) \propto \sum_{E'}|\psi_E'(\mathbf{r})|^2\delta(E-E')$ at $E = 0$ in Fig.~\ref{fig:ldos} a). This clearly confirms the appearance of edge states in the topological regime. 

\subsection{Random lattice}

Recently, it was understood that robust topological states may persist even in systems with little or no spatial order \cite{shenoy,mitchell,poyh2}. As opposed to disordered systems that still possess residual lattice symmetries, spatial order does not play a role in amorphous systems with randomly distributed lattice sites. The fact that amorphous systems can support topologically non-trivial states is truly a manifestation of the independence of spatial symmetries and topology. Here we show that essentially the same topological state engineering that was described in the square lattice with magnetic impurities can be generalized to randomly localized impurities.

\begin{figure}[t]
\centering
\includegraphics[width=0.95\linewidth]{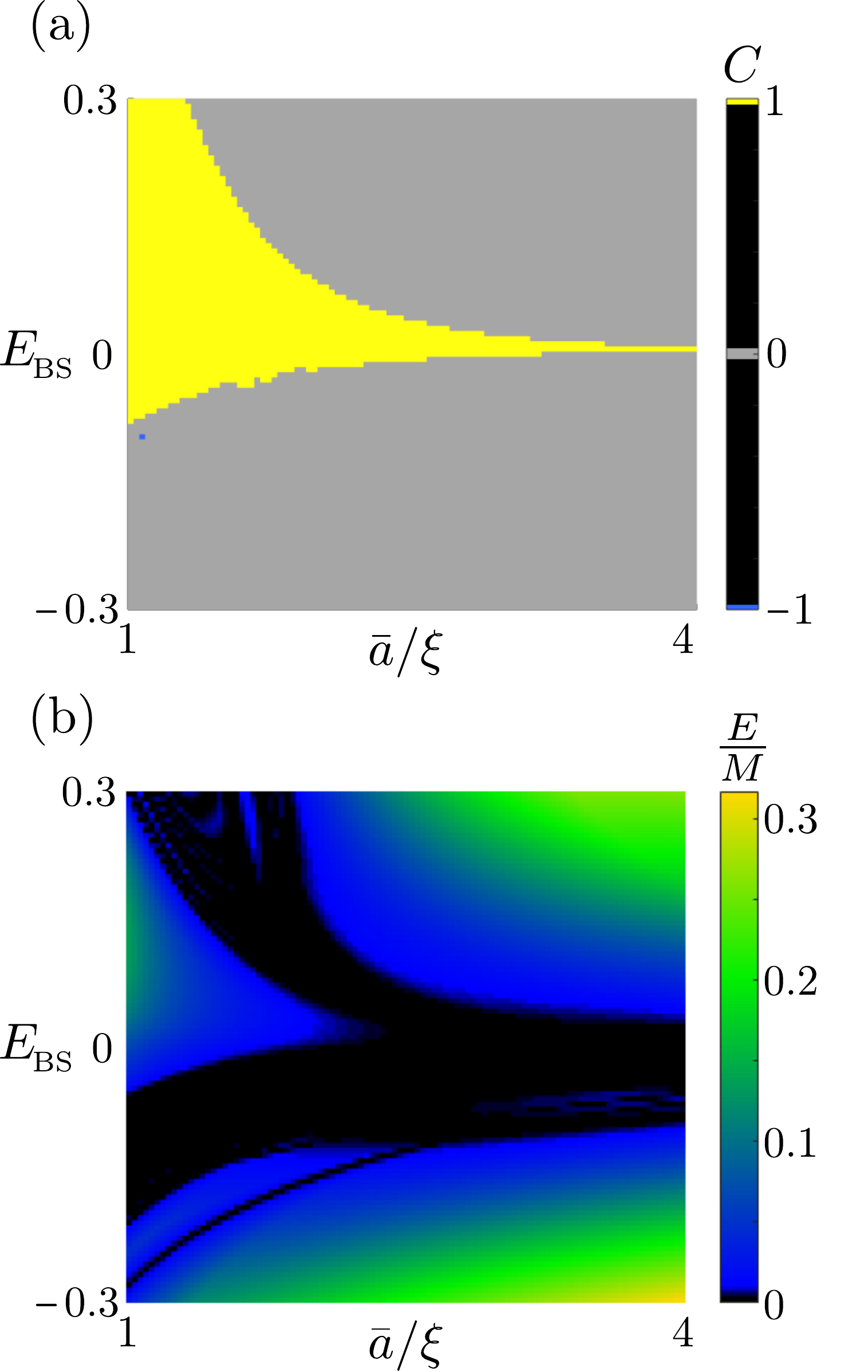}
\caption{a): Topological phase diagram of a single realization of a finite random lattice with periodic boundary conditions.  b): The energy gap diagram of corresponding to the system in Fig.~a) Calculated for a 400 impurities, with $g = 0.15$. }\label{fig:rand}
\end{figure}

The Chern number in translationally non-invariant systems can be defined by studying the response of the ground state to the boundary conditions \cite{niu}. Practical tools to calculate topological numbers is provided by real-space formulas \cite{zhang2,loring}. By employing the method of Ref.~\onlinecite{zhang2}, we have evaluated the topological phase diagram in terms of the characteristic distance between impurities defined as $\bar{a}=1/\sqrt{\rho}$, where $\rho$ is the number of impurities per unit area. As shown in Fig.~\ref{fig:rand}, finite random systems exhibit basically similar topological behaviour as the square lattice.  Also, topological phase diagrams for finite random systems are self-averaging in the sense that differences between different lattice realizations become small already for moderate system sizes. The topological edge modes depicted in Fig.~\ref{fig:ldos} b) are more irregular compared to the square lattice. Nevertheless, the random edge modes support the same quantized transport properties as their counterparts in regular lattices.      

\section{Discussion}\label{sec:engset}
 A goal of this work was to introduce a simple and experimentally motivated model for designer Chern insulators allowing flexible topological state engineering. We adopted a model of a magnetically gapped topological insulator surface decorated by local impurities.  We showed that it is possible to change the topological state of the system by employing magnetic impurities that locally reduce the magnetization of the uniform background.  In fact, the impurities can be regarded as a spatially modulated reduction of the uniform magnetic background. This interpretation suggests  another method to realize the studied topological manipulation.  While the experimental realization of our model remains challenging at the moment, the remarkable recent progress in magnetically doped topological insulators\cite{chang,chang2,bestwick} suggests that patterned magnetization could be accessible in the near future \cite{yasuda}. 
 
The underlying homogeneously magnetized topological insulator surface without impurities is itself topologically non-trivial, exhibiting a quantum anomalous Hall effect\cite{haldane} with half-quantized Hall conductance $\pm e^2/(2h)$ . However, the impurity-based topological state manipulation offers key advantages in fabricating novel functional structures. When crossing the interface between a clean area and an impurity lattice, the value of the Chern number jumps by unity. The impurity lattice thus boosts the magnitude of the Chern number  from $1/2$ to $3/2$ per surface. Taking into account that the sign of the perpendicular magnetization could be changed, the studied system possesses four non-trivial topological states with Chern numbers $\pm 1/2$ and $\pm 3/2$ per surface. Furthermore, the crucial advantage of the impurity lattice is that it allows the tuning of a topological state by electrical means without need to flip the magnetization. The impurity patterning gives rise to two subgap bands $E_k$ depicted in Fig.~\ref{fig:sketch} b). By tuning the electron density of the system by a gate voltage, it is possible to tune the Fermi level with respect to the two subgap bands. The Chern number jumps by unity if the Fermi level is moved outside the band gap of a topologically non-trivial impurity band. Local gates would allow spatially resolved modulation of the Fermi level and thus, in principle, a local tuning of a topological state.  

\begin{figure}[b]
\centering
\includegraphics[width=0.9\columnwidth]{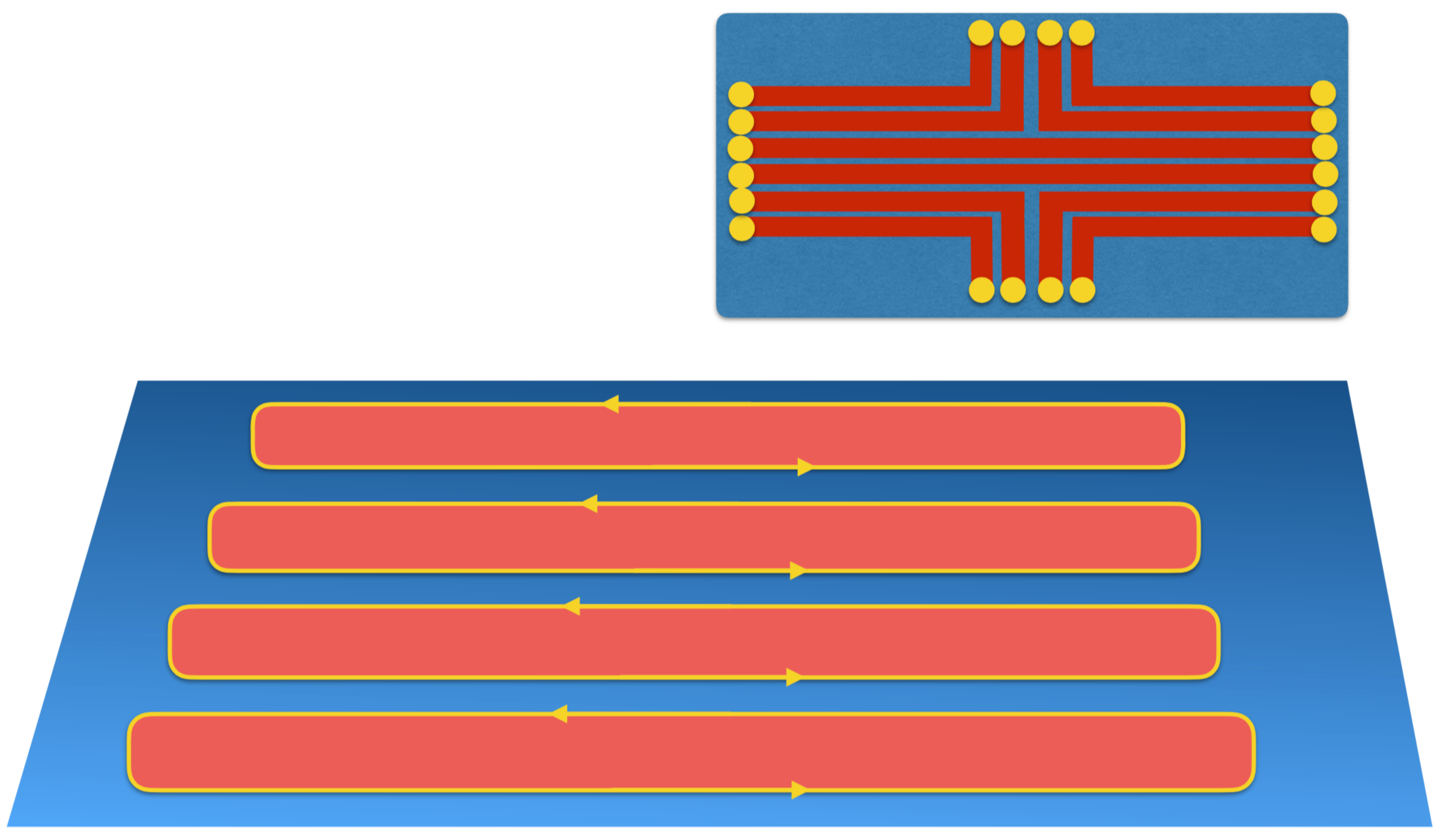}
\caption{On impurity-doped areas (red) the Chern number jump by unity compared to the underlying substrate. The difference in the topological state is accompanied by a chiral edge mode circulating the impurity-rich areas. Inset: By patterning the substrate with impurities it is possible to create edge state circuits that connect different terminals (yellow) on a chip. }\label{fig:edge}
\end{figure}

As dictated by the bulk-boundary correspondence and discussed above, a jump in the Chern number is accompanied by the appearance of a topologically-protected chiral edge mode separating the impurity lattice from clean regions. This is illustrated in Fig.~\ref{fig:edge} where red regions on the blue bulk substrate represent impurity-doped areas. 
Chiral edge modes act as ideal 1D electric channels where backscattering of electrons is completely suppressed. Each channel supports a quantized conductance $e^2/h$ when electron-phonon processes can be neglected. These properties are highly desirable in electronic applications and their successful implementation would be a step towards a dramatic reduction of power loss in electronic devices. A controlled on-chip patterning of topological regions on a substrate would open a door for harnessing the lossless edge mode transport.  As illustrated in Fig.~\ref{fig:edge}, by controlled impurity doping it is possible to imprint multiple topological domains with accompanying edge modes. 

 In Fig.~\ref{fig:edge} (inset), we envision an integrated topological circuit which can be flexibly imprinted on a substrate and where edge modes connect different terminals. As mentioned above, the topological areas could also be defined by locally tuning the filling by electrical means. The edge modes would allow lossless flow of currents between terminals and combine into larger functional units. Edge state circuits could also serve as thermal guides capable of removing parasitic heat from chosen terminals.  

\section{Conclusion}\label{sec:conclusion}

In this work we studied a possibility of engineering designer Chern insulators through impurity doping. Our work established a method of boosting the functionality of the quantum anomalous Hall effect in magnetic topological insulators. The magnetic patterning can regardless of magnet configuration lead to robust topological domains where the Chern number is increased by one compared to the half-quantized background value. More generally,  our work paves the way for on-chip fabrication of tunable Chern insulators and harnessing the unique properties of topological matter in applications. 
We note that an alternative scheme to engineer a Chern insulator starting from a thin film strong topological insulator has already been realized in an experiment~\cite{cherninsulator2013}. The mechanism used is not based on bound states but on the surface properties alone. 

An interesting question for future research is how to fabricate Chern insulators by impurities on topologically trivial platforms. In addition, realizing robust states with higher Chern numbers as in superconductors and achieving topological states by potential impurities would further advance the field of designer topological systems.

{\it {Acknowledgement:}}
E.M. acknowledges funding from the Irish Research Council Enterprise Partnership Scheme, through grant EPSPG/2017/343. This work is part of the D-ITP consortium, a program of the Netherlands Organisation for Scientific Research (NWO) that is funded by the Dutch Ministry of Education, Culture and Science (OCW). K.P. acknowledges the Swedish Cultural Foundation in Finland for support.

E. L. Minarelli, K. P\"oyh\"onen, and G. A. R. van Dalum contributed equally to this work.

\begin{widetext}
\appendix

\section{Expressions for the Green functions}
In this section, we will prove Eqs. \eqref{eq:FullGreen} and \eqref{Eq:SingleGreen} from the main text by solving the integrals
\begin{align}
\hat{\mathcal{G}}_0^+(\mathbf r;E) &= \int \frac{\mathrm{d}^2\mathbf k}{(2\pi)^2}((E+i0^+)\mathbb{1} - \hat{H}_S(\mathbf{k}))^{-1}e^{-i\mathbf k \cdot \mathbf r}\;,\\
\hat{\mathcal{G}}_0^+(0;E) &= \int \frac{\mathrm{d}^2\mathbf k}{(2\pi)^2}((E+i0^+)\mathbb{1} - \hat{H}_S(\mathbf{k}))^{-1}
\end{align}
in terms of the bulk Hamiltonian
\begin{equation}
\hat{H}_S(\mathbf{k}) = v_F(k_x\tau_x + k_y\tau_y) + M\tau_z\;,
\end{equation}
where, as described in the main text, we have set $\mu$ to 0.

The Green function for position $\mathbf r \neq 0$ is, if we transform to radial $k$-space coordinates, given by
\begin{equation}
\hat{\mathcal{G}}_0^+(\mathbf r;E) = \int \frac{\mathrm{d}^2\mathbf k}{(2\pi)^2}\frac{E\mathbb{1} + v_F k(\cos\theta_k\tau_x + \sin\theta_k\tau_y) + M\tau_z}{E^2 - v_F^2k^2 - M^2}e^{-i k r \cos(\theta_k - \theta_r)}\;,
\end{equation}
where $\theta_k$ ($\theta_r$) is the angle between $\mathbf k$ ($\mathbf r$) and the $x$ axis. By looking at the matrix structure, we see that the result will take the form
\begin{equation}
\hat{\mathcal{G}}_0^+(\mathbf r;E) =
\begin{pmatrix}
(E + M)I_1(\mathbf r) & I_2^-(\mathbf r)\\
I_2^+(\mathbf r) & (E-M) I_1(\mathbf r)
\end{pmatrix}\label{eq:gasamatrix}\;,
\end{equation}
where
\begin{align}
I_1(\mathbf r;E) &= \int \frac{\mathrm{d}^2\mathbf k}{(2\pi)^2}\frac{1}{E^2 - v_F^2k^2 - M^2}e^{-i k r \cos(\theta_k - \theta_r)}\;,\\
I_2^\pm(\mathbf r;E) &= \int \frac{\mathrm{d}^2\mathbf k}{(2\pi)^2}\frac{v_F k(\cos\theta_k\pm i\sin\theta_k)}{E^2 - v_F^2k^2 - M^2}e^{-i k r \cos(\theta_k - \theta_r)}\;.
\end{align}
In order to solve these integrals, we will need make use of the following integral equations for, respectively, the Bessel function of the first kind $J_n(x)$ and for the modified Bessel function of the second kind $K_n(x)$:
\begin{align}
\int_0^{2\pi} \mathrm{d}\theta e^{i(x\cos\theta + n\theta)} &= 2\pi i^n J_n(x)\;,\\
\int_0^\infty \mathrm{d}k \frac{k J_0(kx)}{k^2 + b^2} &= K_0(bx)\;,
\end{align} 
where $x\in\mathbb{R}^+$, $n\in\mathbb{Z}$, and $\Re[b]>0$. By straightforward application of the above identities together with the following identities concerning the derivative of the Bessel function with respect to its argument:
\begin{equation}
\begin{aligned}
\dfrac{\partial}{\partial x}J_{0}(x) = - J_{1}(x) &&,&&  \dfrac{\partial}{\partial x}K_{0}(x) = - K_{1}(x)\;,
\end{aligned}
\end{equation}
we find
\begin{align}
I_1(\mathbf r;E) &= - \frac{1}{2\pi v_F^2} K_0\left(\frac{r}{\xi}\right)\;,
\end{align}
where we introduced the correlation length $\xi \equiv v_F/\sqrt{M^2 - E^2}$.
The result is only valid for energies $|E| < |M|$, but as we are focusing on subgap bands, this is true by definition and does not present a problem. For the second integral, we get
\begin{equation}
I_2^\pm(\mathbf r) = \frac{e^{\pm i \theta_r}}{(2\pi)^2}\int_0^\infty \mathrm{d}k  \frac{v_F k^2 }{E^2 - v_F^2k^2 - M^2} \int_0^{2\pi} \mathrm{d}\phi e^{-i k r \cos\phi \pm i\phi}\;,
\end{equation}
with $\phi\equiv\theta_k-\theta_r$. Applying the previously-mentioned Bessel identities, this yields
\begin{equation}
I_2^\pm(\mathbf r) = \frac{i e^{\pm i \theta_r}}{2\pi v_F^2}\sqrt{M^2 - E^2} K_1\left(\frac{r}{\xi}\right)\;.
\end{equation}
Noting that $
e^{\pm i\theta_r} = \cos\theta_r \pm i \sin\theta_r = \frac{x\pm iy}{r} $,
and combining our results into Eq.~\eqref{eq:gasamatrix}, we find that the Green function at nonvanishing $r=|\mathbf r|$ is
\begin{equation}
\hat{\mathcal{G}}^+_0(\mathbf r;E) = 
-\frac{1}{2\pi v_F^2}
\begin{pmatrix}
(E + M)K_0\left(r/\xi\right) & -i \frac{v_F}{\xi}\frac{x-iy}{r}K_1\left(r/\xi\right)\\
-i \frac{v_F}{\xi}\frac{x+iy}{r}K_1\left(r/\xi\right) & (E - M)K_0\left(r/\xi\right)
\end{pmatrix}\;,
\end{equation}
which, when written in terms of Pauli matrices, is Eq.~\eqref{eq:FullGreen} in the main text.

For $\mathbf{r}=0$ the above expression diverges, so it is necessary to consider this case separately:
\begin{equation}
\hat{\mathcal{G}}_0^+(0;E) = \int \frac{\mathrm{d}k_x \mathrm{d}k_y}{(2\pi)^2}\frac{E\mathbb{1} + v_F(k_x\tau_x + k_y\tau_y) + M\tau_z}{E^2 - v_F^2k^2 - M^2}\;.
\end{equation}
The terms linear in $k_x,k_y$ are odd and will therefore vanish, so with the substitution $u = v_F^2 k^2$, we get
\begin{equation}
\hat{\mathcal{G}}_0^+(0;E) = \frac{E\mathbb{1} + M\tau_z}{4\pi v_F^2} \int_0^\infty \frac{\mathrm{d}u}{E^2 - u - M^2}\;.
\end{equation}
The integral diverges as written. We assume instead that it is only well defined up to some bulk bandwidth $D_{bulk}=v_Fk_{max}$, such that
\begin{equation}
\begin{split}
\hat{\mathcal{G}}_0^+(0;E) &\approx \frac{E\mathbb{1} + M\tau_z}{4\pi v_F^2} \int_0^{D_{bulk}^2} \frac{\mathrm{d}u}{E^2 - u - M^2}\\
&= -\frac{E\mathbb{1} + M\tau_z}{4\pi v_F^2} \log\left(1 + \frac{D_{bulk}^2}{M^2 - E^2}\right)\;,
\end{split}
\end{equation}
which corresponds to Eq.~\eqref{Eq:SingleGreen}. If $D_{bulk} \gg M$ and $M \gg E$, this reduces to
\begin{equation}
\hat{\mathcal{G}}_0^+(0;E) \approx -\frac{E\mathbb{1} + M\tau_z}{2\pi v_F^2} \log\left(\frac{D_{bulk}}{|M|}\right)\;.
\end{equation}

\section{Derivation of the tight-binding model}
The full Hamiltonian of the system is of the form
\begin{equation}
\hat{H} = \hat{H}_0 + \delta\hat{H}(\mathbf r)\;,
\end{equation}
where $\hat{H}_0=\hat{H}_S$ is translationally invariant, and the spatially dependent part can be described by
\begin{align}
\delta\hat{H}(\mathbf{r})=\hat{H}_I(\mathbf{r})= \sum_j \hat V(\mathbf{r}) \delta(\mathbf r - \mathbf r_j)= \sum_j (V_{0} \mathbb{1}+V_{M}\tau_{z}) \delta(\textbf{r}-\textbf{r}_{j})
\end{align}
for a collection of pointlike impurities at sites $\lbrace\mathbf r_j\rbrace$.
From the Schrödinger equation $\hat{H}\psi = E\psi$ we get, by grouping the translationally symmetric  terms to one side,
\begin{equation}
\big(E-\hat{H}_0\big)\psi(\mathbf r) = \sum_j\hat V(\mathbf r)\delta(\mathbf r - \mathbf r_j)\psi(\mathbf r)\;.
\end{equation}
A Fourier transform yields
\begin{equation}
\big(E - \hat{H}_0(\mathbf k)\big)\psi(\mathbf k) = \sum_j \hat V(\mathbf r_j) \psi(\mathbf r_j)e^{i\mathbf k \cdot \mathbf r_j}\;.
\end{equation}
We can then multiply by $e^{-i \mathbf k \cdot \mathbf r_i}$ for some $\mathbf r_i \in \lbrace \mathbf r_j\rbrace$ and integrate to get
\begin{equation}
\psi(\mathbf r_i) = \sum_j \left(\int \frac{\mathrm{d}\mathbf k}{(2\pi)^2} \hat{\mathcal{G}}_0(\mathbf k;E) e^{i\mathbf k \cdot (\mathbf r_j - \mathbf r_i)} \right) \hat V(\mathbf r_j) \psi(\mathbf r_j)\;,
\end{equation}
where $\hat{\mathcal{G}}_0(\mathbf{k};E)$ is the Green function corresponding to the bulk Hamiltonian $\hat{H}_0$. This can be regrouped into Eq.~\eqref{eq:mi} of the main text by separating the terms at $\mathbf r_i$.

\end{widetext}

\end{document}